%% file: main.tex
\documentclass[conference]{IEEEtran}
\IEEEoverridecommandlockouts
\usepackage{cite}
\usepackage{amsmath,amssymb,amsfonts}
\usepackage{algorithmic}
\usepackage{graphicx}
\usepackage{textcomp}
\usepackage{xcolor}
\usepackage[utf8]{inputenc}

\usepackage[hyphens]{url}
\usepackage{hyperref}

\usepackage{enumerate}

\newcommand{\lmttfont}{\fontfamily{lmtt}\selectfont}

\def\BibTeX{{\rm B\kern-.05em{\sc i\kern-.025em b}\kern-.08em
    T\kern-.1667em\lower.7ex\hbox{E}\kern-.125emX}}

\begin{document}

\title{Isolating Real-Time Safety-Critical Embedded Systems via SGX-based Lightweight Virtualization}


\author{

\IEEEauthorblockN{Luigi De Simone}
\IEEEauthorblockA{Università degli Studi di Napoli Federico II, Italy\\
luigi.desimone@unina.it}\\
\and
\IEEEauthorblockN{Giovanni Mazzeo}
\IEEEauthorblockA{University of Naples ``Parthenope'', Italy\\
giovanni.mazzeo@uniparthenope.it}

}

\maketitle
\thispagestyle{plain}
\pagestyle{plain}

\begin{abstract}
A promising approach for designing critical embedded systems is based on virtualization technologies and multi-core platforms. These enable the deployment of both real-time and general-purpose systems with different criticalities in a single host. Integrating virtualization while also meeting the real-time and isolation requirements is non-trivial, and poses significant challenges especially in terms of certification. In recent years, researchers proposed hardware-assisted solutions to face issues coming from virtualization, and recently the use of \emph{Operating System (OS) virtualization} as a more lightweight approach. Industries are hampered in leveraging this latter type of virtualization despite the clear benefits it introduces, such as reduced overhead, higher scalability, and effortless certification since there is still lack of approaches to address drawbacks.
In this position paper, we propose the usage of Intel's CPU security extension, namely SGX, to enable the adoption of enclaves based on \emph{unikernel}, a flavor of OS-level virtualization, in the context of real-time systems. We present the advantages of leveraging both the SGX isolation and the unikernel features in order to meet the requirements of safety-critical real-time systems and ease the certification process.
\end{abstract}
\begin{IEEEkeywords}
Real-time, Intel SGX, Unikernel, Virtualization
\end{IEEEkeywords}

\section{Introduction}
\input{introduction.tex}

\section{OS-level Virtualization in real-time domain}
\input{towards_os_level_real_time.tex}

\section{Enhancing Isolation via Intel SGX}
\input{sgx-isolation.tex}

\section{Certification implications}
\input{certification_issues.tex}

\section{Conclusion}
\input{conclusion.tex}

\section*{Acknowledgment}

This work has been founded by RFI - Rete Ferroviaria Italiana - within the Research Project “Analisi ed implementazioni prototipali di funzioni di interfaccia tra kernel e applicativi software di logiche di segnalamento”, Grant reference number: DAC n. 552/2017– CIG 710174515F.


\IEEEtriggeratref{50}

\bibliographystyle{IEEEtran}
\bibliography{bibliography}

\end{document}

%% file: introduction.tex
In recent years, \emph{Critical Real-Time Embedded Systems} (CRTES) are shifting towards an \textit{integrated paradigm}, in which multiple applications share the same physical platform resources \cite{di2010moving}. Such an approach allows running applications with different level of criticality on the same embedded platform. The enabling technologies have been multi-core processors and virtualization. The former enhances performance by reducing the overall costs, size, weight, and energy consumption \cite{burns2018survey}. While the latter is typically adopted to create separate domains of CRTES tasks. Despite promises given by these technologies, there are several drawbacks to be addressed \cite{garcia2014challenges}. More precisely, the temporal, spatial, fault, and I/O isolations between virtual domains are at risk in hypervisors' hands. Additionally, the design of CRTESs is still bounded to a rigorous certification process, in which developers must provide evidence about a huge amount of software and its level of partitioning, showing documented proofs about, e.g., fault containment between a virtual domain to another, absence of temporal interferences between critical and non-critical functions. In order to increase the isolation of virtual domains, the research community explored the possibility of leveraging \emph{hardware-assisted} solutions. The most explored approach has been to use security features provided by ARM, i.e., the ARM TrustZone technology \cite{trustzone}, indeed enabling a more powerful form of isolation assisted by the hardware. ARM TrustZone, in fact, supports a so-called ``dual world'' execution \cite{cereia2008asymmetric, douglas2010thin, frenzel2010arm, wilson2007implementing, pinto2017ltzvisor, winter2008trusted, cereia2009virtual, sangorrin2010dual, oh2012acceleration, pinto2014towards, schwarz2014affordable}. Usually, the \textit{non-secure world} is used as an environment for running VMs, which are managed by the hypervisor software that runs in the \textit{secure world}. Researchers used the TrustZone dual-guest configuration for running a general-purpose OS (GPOS) within the non-secure world, while a real-time OS (RTOS) runs in the secure-world having a full view of the entire system. In this way, the critical tasks running on top of the RTOS are isolated by non-critical tasks.
\\Another emerging trend in real-time applications is exploiting \emph{OS-level virtualization} (also named container-based virtualization) \cite{rtcase}, which ---unlike fully emulating a hardware machine--- it abstracts OS processes (called containers) by extending the (host) OS kernel. The main reason behind the usage of containers in real-time domains \cite{kubernetes_frakti, clear_container, kata_container, hyper_container} is to reduce the overhead affecting VMs and better scale when a larger number of applications of different criticalities are in place. Indeed, the container approach does not require to replicate the entire OS environment for every system. However, the performance benefits introduced by containers come at the cost of reduced isolation, threatening the practicability of OS-level virtualization under real-time and safety requirements. In the context of OS-level virtualization, the sole approach to somehow face isolation issues may be to run a single application in its virtual domain. Such model is known as \textit{unikernel} or library OSes \cite{unikernel}, in which the full software stack of a system, including OS components, libraries, language runtime, and applications, is compiled into a single VM that runs directly on a general-purpose hypervisor. This approach introduces benefits such a small code base, low attack surface, and an effortless certification process due to the low amount of software to be verified \cite{unikernel}. However, stronger isolation proofs to be reported to certifiers are still lacking. 

In this position paper, we propose a hardware-assisted solution for leveraging \emph{OS} virtualization in the field of CRTES. 
In particular, we discuss the potential adoption of the well established Intel \emph{Software Guard eXtension} (SGX) extension \cite{intel_sgx} to enable powerful isolation when OS-level virtualization is pursued in CRTES applications having certification requirements. The SGX is the technological implementation of hardware-assisted trusted execution, as conceived from Intel. SGX belongs to the same category of ARM TrustZone since its \emph{Trusted Execution Environment} (TEE) is internal to the CPU perimeter. Intel SGX provides capabilities for securing user space applications without the need of calling privileged OS code. Basically, SGX provides tools for creating memory areas called \textit{enclaves}, which protect application code and data from accesses by other software, including higher-privileged software. Memory pages which are within an enclave can not be accessed by code outside of the enclave. 
The proposed approach is intended for executing an RTOS based on \emph{unikernels} into an SGX enclave. Besides the advantages for CRTES, unikernels represent also one of the best ways to bypass the impossibility of issuing \emph{syscalls} (normally executed at \emph{ring0}) from within an enclave that only supports \emph{ring3} functions \cite{arnautov, sgxlkl, tsai, haven}. Unfortunately, nowadays, the availability of unikernel-based RTOSes are still poor. \emph{MirageOS} \cite{mirageos} is the sole example. 
Finally, we examine the impact that such an approach could entail during the certification process, by focusing on the isolation properties that must be fulfilled.




%% file: towards_os_level_real_time.tex
Virtualization technologies are the core enabler of several computer engineering applications, ranging from cloud computing to real-time embedded systems. In contrast to cloud computing, in CRTES development there is a need for specific mechanisms that guarantee and enforce the execution of applications to meet timing and safety requirements \cite{garcia2014challenges}. In years, several techniques were developed to abstract physical resources in virtual, from the classical full-virtualization and para-virtualization, to more recent \textit{OS-level} virtualization \cite{Tanenbaum:2014:MOS:2655363, vanMoolenbroek:2014:TFL:2611354.2611369}.


\textit{OS-level} virtualization, allows running multiple appliances without hardware virtualization. The idea behind container-based virtualization is to enhance the abstractions of OS processes (called containers), by extending the (host) OS kernel, in order to have a virtual domain with its own virtual CPU and virtual memory like in traditional OS processes, a virtual filesystem, virtual network, IPCs, PIDs, and users management. These virtual resources are distinct for each container in the system.
Currently, the most used container-based virtualization technologies are LXC \cite{lxc}, Docker \cite{docker}, and OpenVZ \cite{openvz}.

As mentioned before, virtualization has to address the critical problem of guarantee the isolation among virtual instances \cite{bugnion2012bringing, inject_hw_fault_hypervisor, jailhouse_perf_isolation_test}. In the more general sense, isolation means the fact that something is independent and disentangled to the behaviors of other things. Thus, the virtualization layer has to be in complete control of virtualized resources, and applications running on a virtual domain must have the illusion to be completely isolated from others.
In general, in the virtualization context, we consider two main isolation properties that are mandatory.

The \emph{temporal isolation} (also known as \textit{performance isolation} or \textit{temporal segregation}), is the capability of isolating or limiting the impact of resource consumption (e.g., CPU, network, disk) of a virtual domain on the performance degradation of the other virtual domains, but also against the host. This means that a critical task running on a virtual domain (e.g., task on a VM or within a container) must not cause severe delays of other critical and non-critical tasks, leading to a phenomenon like starvation, reduction of throughput, and increased latency. Temporal isolation is crucial in embedded systems when critical tasks within containers need to assure SLA guarantees about performance. In the context of safety-critical applications, some standards (e.g., IEC 61508-3 Annex F \cite{iec61508}, ISO 26262-6 Annex-D \cite{iso26262}, ARINC-653 \cite{arinc653}, DO 178 6.3.3f \cite{do178b}, CAST-32A \cite{CAST-32A}) suggest to exploit a cyclic scheduling among virtual domains. 


The other crucial isolation property is the \emph{spatial isolation} (also known as \textit{memory isolation} or \textit{spatial separation}). Such property describes the capability of isolating code and data between virtual domains, and between virtual domains and the host kernel. This means that a task should not be able to alter private data belonging to other tasks, including the devices allocated to a specific task. Break spatial isolation will likely lead to unexpected behaviors or worse a system crash \cite{break_virtual_mem_sgx}.
Usually, spatial isolation is enforced by hardware memory protection like Memory Management Unit (MMU), which protects the task's virtual memory space. Furthermore, by considering the case of shared devices, we need to focus also on the \emph{I/O isolation} property. Often, the IOMMU is exploited to properly address the isolation of memory-mapped devices, and in some cases, the access on hardware devices from a different virtual domain is serialized.

Finally, another kind of isolation property that must be addressed is \emph{fault isolation} (also known as \textit{fault/error containment}). Fault isolation means that potential faults that occur in one virtual domain should not be propagated towards hypervisor and/or other partitions, leading to hangs or even halting the entire system.


Bearing in mind the needs of a virtualization approach that is both lightweight and provide the isolation properties mentioned above, \emph{unikernels} are recently gaining a significant attention \cite{mikelangelo}. The \textit{unikernel} is an approach of linking an application with OS components. In particular, such components include the core services of a kernel like memory management, scheduler, network and disk stack, and device drivers. Thus, the application and the kernel have a unique address space, creating a standalone binary image that is bootable directly on physical and/or virtual hardware \cite{madhavapeddy2013unikernels}.
The big advantage that unikernels provide is that the kernel functionalities can be specialized to fit the needs of the target application. For example, the developer would want to increase a specific aspect of performance of its application (e.g., the network throughput) or to improve isolation.
Since unikernel is a lightweight solution for virtualization, it could be a promising solution to be adopted in the context of safety-critical embedded systems, where the needs of having a well-defined software component would facilitate the certification process.
Another example of the potential of unikernel in the embedded systems is the predictability. Indeed, since moment by moment, only one task/process is running on the unikernel, we can say with good confidence that if an operation is completed in a specific amount of time, it will take the same amount of time every time will be executed. If we compare such behavior against the Linux kernel, for example, there are various factors of unpredictability, like page faults, internal locks, scheduling jitter, and so on. Thus, unikernels would be also a promising solution for the measurement of the worst-case execution time (WCET), which is mandatory and usually is a non-trivial task both in single- and multi-core platforms. Naturally, we need to consider that the predictability of unikernels is valid in the real-time domain as long as the underlying hypervisor schedules the task to the CPU always in the same way. Thus, we need to take into account the possibility of using real-time hypervisors for solving the problem of determinism.

\begin{figure}[h]
    \centering
    \includegraphics [scale=0.22] {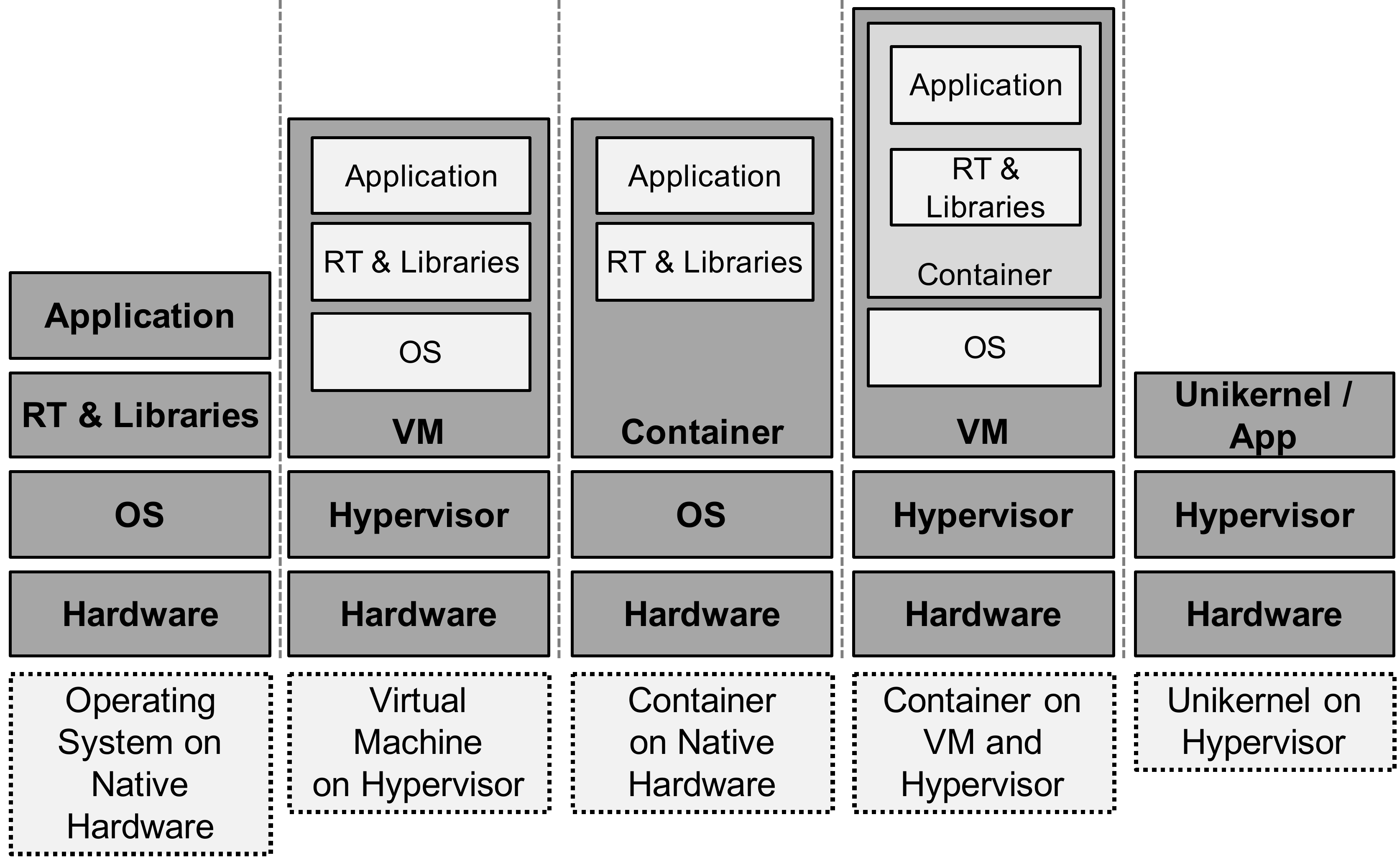}
    \caption{Virtualization deployments.}
    \label{fig:virtualization_examples}
\end{figure}

\figurename{}~\ref{fig:virtualization_examples} summarizes potential virtualization approaches that could be used in the development of CRTES, from the classical hypervisor-based, passing through the container-based to the unikernel model.


    
%
%
%

%% file: sgx-isolation.tex
\label{sgx-isolation}

\begin{figure}[h]
    \centering
    \includegraphics [scale=0.49] {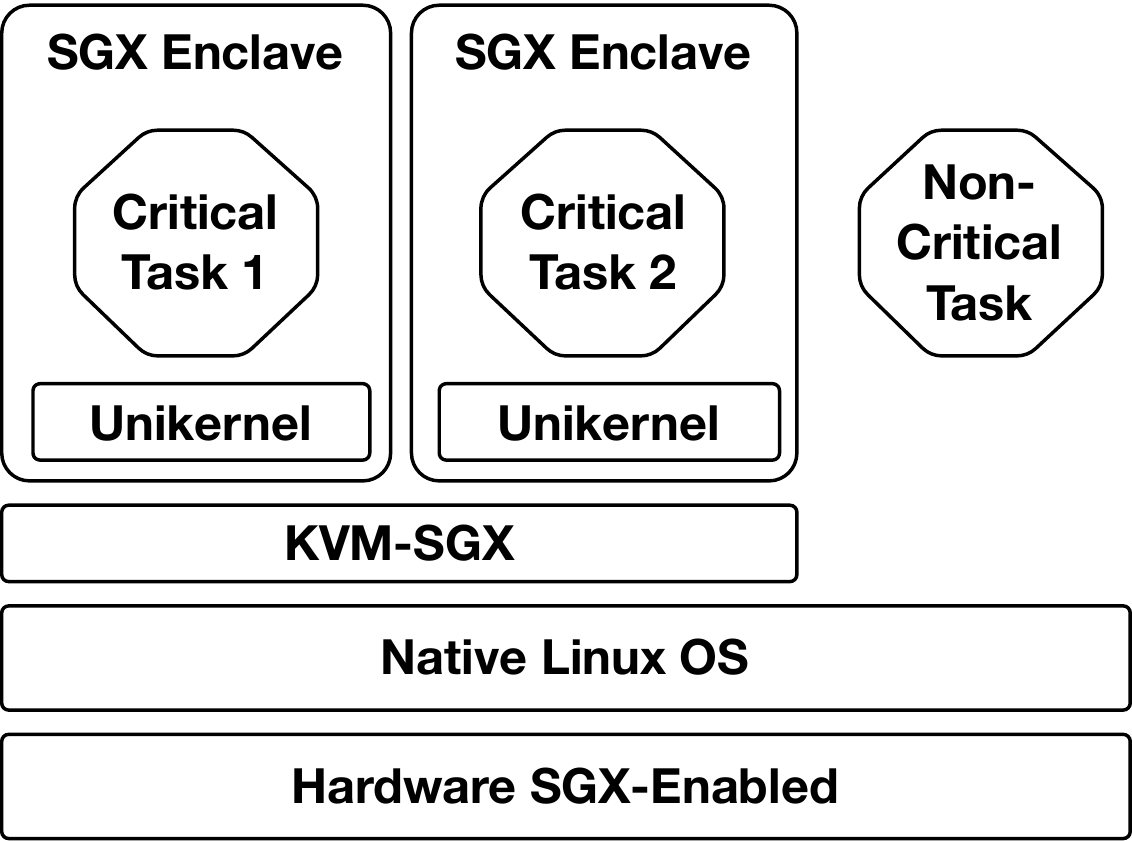}
    \caption{Architecture of CRTES based on Intel SGX}
    \label{fig:arch}
\end{figure}

Leveraging OS virtualization for CRTES is non-trivial as the confinement of containers' domains is more difficult to enforce and demonstrate, due to the weak isolation that LXC \emph{namespace} and \emph{cgroup} create \cite{nfv_bench}.
In this sense, a \emph{unikernel} could represent a solution. Its properties of an extra lightweight OS can certainly facilitate the certification process in terms of software verification. However, \emph{unikernels} are still subjected to isolation problems that could lead to interferences among critical tasks of CRTES. In this paper, we propose the adoption of Intel SGX-enabled \emph{unikernels} in the context of CRTES to provide guarantees on the isolation for real-time tasks running in dedicated domains. In fact, the hardware-assisted security isolation features of Intel SGX can be leveraged to this end as other researchers similarly did with ARM TrustZone to segregate VM domains \cite{cereia2008asymmetric, douglas2010thin, frenzel2010arm, wilson2007implementing, pinto2017ltzvisor, winter2008trusted, cereia2009virtual, sangorrin2010dual, oh2012acceleration, pinto2014towards, schwarz2014affordable}. 
\\SGX is a security technology that is catching on in both research and industrial communities. It is a set of new hardware instructions that isolate sensitive code and data processing, even from users with \emph{root} privileges. SGX enables a confined region of execution, namely \emph{secure enclave}, whose access is supervised by the hardware that enforces \emph{isolation} of processes. The SGX threat model assumes that most of the host stack is untrusted. Thus, the CPU ensures that the enclave memory is not accessible by any part of the system, except for the code running inside the enclave itself. For the purpose of this paper, it is important to notice that there are two main drawbacks of SGX to be taken into account. That is, \emph{i}) the physical memory size dedicated to all the instantiated enclaves is limited to $128MB$, \emph{ii}) the execution of \emph{syscalls} is forbidden in the enclave as the OS is considered untrusted.
\\Currently, there are solutions where SGX was used to enhance the security of unmodified applications basically by running them in SGX-secured domains \cite{arnautov, sgxlkl, tsai, haven}. These studies pursued two different approaches for providing \emph{syscall} support to the application into an enclave \cite{comparative}. In particular, in a first case Arnautov et al. \cite{arnautov} use an SGX-extended \emph{libc} library to expose an external (and optimized) \emph{syscall} interface, which is SGX-shielded. In a second case \cite{sgxlkl, tsai, haven}, researchers leverage the single-address space property of \emph{unikernels} (i.e., \emph{Linux-Kernel-Library}, \emph{Graphene}, and \emph{Drawbridge}, respectively) to execute \emph{syscalls} directly inside the SGX enclave. Essentially, the latter category of studies port an entire \emph{unikernel} into an enclave to provide lightweight OS support to the application. Contrariwise, Sfyrakis et al. \cite{uniguard} propose the adoption of SGX to secure only some computations of a \emph{MirageOS} \emph{unikernel} \cite{mirageos}.
\\In this position paper, we lay the foundations for a solution where \emph{unikernels} ---of different or same typology--- run tasks of distinct criticality in SGX enclaves. Figure \ref{fig:arch} shows a possible architecture of our proposal. 

The proposed approach is intended for only running critical tasks in SGX and the non-critical in the ``normal'' world. Another approach would have been to launch all the tasks in as many enclaves. However, unlike solutions based on ARM TrustZone, we need to take into account the memory limitation imposed by SGX technology, i.e., $128MB$, which is shared by all the enclaves. Hence, we believe that one enclave per critical task is the only feasible approach. Even in this case, launching too many critical tasks could entail that the memory bounds would be exceeded. A more accurate number highly depends on the \emph{unikernel} solution adopted. For example, \emph{IncludeOS} is a unikernel with one of the lowest memory footprint (i.e., $\approx 10MB$). Currently, \emph{unikernels} footprints are approximately of $8MB$ on average \cite{morabito2018consolidate}. This means that we could have a number of about $16$ running critical tasks, which are in many cases above the real need for a CRTES.
\\Moreover, the adoption of SGX for CRTES can not disregard an analysis on the \emph{isolation} properties. The \emph{spatial} and \emph{temporal isolation} are essentially ensured by SGX. The CPU realizes architectural isolation by monitoring the accesses to enclave-owned physical memory pages. The mechanism of enclave isolation is mainly provided through the \emph{Enclave Page Cache Map} (EPCM). 
The EPCM is the table where the SGX processor checks the correctness of the system software’s allocation decisions, and refuse to perform any action that would compromise SGX isolation. An EPCM entry identifies the enclave that owns the \emph{Enclave Page Cache} (EPC) page, that is, the enclaves' content and the related data structures information. Since the EPCM identifies one specific owning enclave for each EPC page, enclaves can not communicate via shared memory using EPC pages. Enclaves only share untrusted non-EPC memory. SGX also prevents attempts from the Direct Memory Access (DMA) devices to get access to the enclave. This is of importance as current memory isolation techniques for CRTES are usually based on the MMU, which prevents applications running in one partition to read/write into address space allocated to other partitions. However, such an isolation \emph{as-is} is threatened by the DMA that could bypass the checking procedure of the MMU.
In a nutshell, the isolation may be put at risk when, e.g., the \emph{unikernel} running in the enclave needs support from the external world or needs to communicate with another critical task executing in another enclave. In fact, there are some \emph{unikernel} \emph{syscalls} that must necessarily demand to the external OS. The implemented solution of an SGX-enabled CRTES should take into account the previous considerations. Solving mechanisms might be designed and developed to face the \emph{isolation} of tasks and provide further guarantees in the certification phase.

%% file: certification_issues.tex


Regardless of the specific domain, the developing of safety-critical applications in industries raises several concerns from the certification point of view. In order to reach a specific level of safety, indicated as Safety Integrity Level (SIL) (but depending on the standards, SIL appears also as Automotive Safety Integrity Level (ASIL), Software Safety Integrity Level (SSIL), Design Assurance Level (DAL)), standards require performing burdensome tasks that include verification, performance testing, impact analysis, use of coding standards, on both the hardware and software components within the developed systems. Notwithstanding the cost and effort of certifying CRTES increases significantly with the SIL level, the problem is more exacerbated due to the integration of commercial off-the-shelf (COTS) hardware and software in the products. In general, compared to the bare-metal solutions, the use of virtualization brings additional software layers and components in the overall architecture, and this lead to further complicate the certification. Thus, providing evidence for isolation properties is far from to be effortless.

Considering the hardware and software stacks in the development process, we need to address the different level of safety required by the certification process, which lead to analyze isolation properties at different layers.
For example, according to the EN 50128 standard that provides guidelines for the certification of the software employed in the railway domain \cite{cenelec201150128, sil2}, focusing on the temporal and spatial isolation guarantees, the requirements {\lmttfont D.45 Response Timing and Memory Constraints} specifies that is needed an analysis aimed to estimate the resource usage and the latency for each system functionalities, which include all software modules (from the hypervisor to the kernel).

The strategy for developing CRTES based on virtualization and multi-core platform should be based on the guidelines provided by standards. 
The safety standards impose a precise development process (e.g., the IEC 61508 recommends using the V-model development process for designing safety-related software and hardware \cite{iec61508}), where we need to comply with further steps and requirements for covering safety and certification needs. Normally, such process should be based on the management of systems hazards, meaning that such hazards are eliminated or at least mitigated enough up to tolerable rates for the safety levels assigned to the identified safety functions. 
A potential approach for identifying the possible threats to safety is an analysis of the failure modes. For example, according to our proposal, we could apply the Failure Mode and Effect Analysis (FMEA), and try to identify the potential causes of failures by introducing disturbances (e.g., faults, anomalous loads) at the different levels of the CRTES software stack.
Actually, many studies in literature \cite{cotroneo2013fault, cotroneo2012experimental, cotroneo2018run, cotroneo2016faultprog, nfv_dep_guidelines, winter2015no} and various international standards for software reliability and safety \cite{iso26262, nasastd2004, do178b, ISO/IEC25045, cenelec201150128} exploit the injection of faults in complex systems in order to assess their behavior and unveil potential bottlenecks and critical components under these abnormal inputs and conditions. Furthermore, fault injection technique is often used to measure the efficiency (e.g., coverage, latency, etc.) of fault tolerance mechanisms, including fault detection and recovery. For example, in the EN 50128 document is clearly stated that for the maximum level of safety (i.e., SIL 4) is highly recommended using Software Error Effect Analysis (SEEA) for identifying the criticality of each software component and improve the overall robustness of the software.

In combination with our proposal, there is a clear need for developing systematic approaches for testing the degree of isolation required by certification. In particular, by leveraging fault injection technique, an approach would be identifying and enumerating all the interfaces and resources involved at different levels of the CRTES software stack (e.g., hypervisor- and unikernel-level), to find suitable locations for injecting faults. Furthermore, it would be necessary defining measures of spatial and temporal isolation that could point out isolation issues; for example, a developer could use the existing performance isolation metrics \cite{perf_isol_metrics}, or adapt other metrics in the context of isolation \cite{hiller2001dataerrors}.


Independently from the above considerations, it is important to underline that safety standards do not oblige developers or practitioners to use some particular measure or procedure for evaluating the fault tolerance, because they are not meant for exactly this or that target system. Instead, they suggest general guidelines that apply to a more wide family of systems.

%% file: conclusion.tex
In this position paper, we have discussed the adoption of both the Intel SGX hardware extension and the unikernel lightweight virtualization in the context of safety-critical real-time embedded systems. Our proposed idea is to leverage the isolation properties provided by SGX ---typically used for security reasons in untrusted systems--- in order to provide \emph{temporal} and \emph{spatial} isolation guarantees for critical tasks executing in the trusted execution environment of SGX. 
Potentially, our work could introduce the following advantages to CRTES:

\begin{enumerate}[i)]

    \item the isolation among critical tasks is stronger thanks to the boundaries checks enforced by SGX hardware; 
    
    \item the reliability increases as the risk of faults (e.g., memory leak) with \emph{unikernel} is much more reduced, which is crucial for safety-related systems;
    
    \item the performance is higher as the \emph{unikernel} model indeed offers performance improvements;
    
    \item the software certification process is facilitated thanks to \emph{unikernels}' lightweight properties.

\end{enumerate}
    
We want to remark that a fundamental requirement is the availability of the SGX hardware. Currently, the ARM architecture is widely used as 32b/64b RISC multi-core processors in embedded systems. However, we are witnessing the trend in the real-time embedded world of adopting virtualization to support multi-purpose OSes. In this regard, the most comprehensive virtualization support comes from Intel's architecture. Hence, the future usage of these CPUs is very likely, as witnessed by Intel itself \cite{intelcpurts}.